\documentstyle[twoside,fleqn,epsfig]{actaps}

\begin{document}

\headings{1}{10}

\def\authorlist{K. Banaszek, K. W\'odkiewicz}
\def\shorttitle{Nonlocality of the EPR state\ldots}

\title{\uppercase{Nonlocality of the Einstein-Podolsky-Rosen state\\
in~the phase space}}

\author{{Konrad Banaszek$^a$}%
\email{Konrad.Banaszek@fuw.edu.pl},
{Krzysztof W\'odkiewicz$^{ab}$}%
\email{Krzysztof.Wodkiewicz@fuw.edu.pl}}{(a) Instytut
Fizyki Teoretycznej, Uniwersytet Warszawski, Ho\.{z}a 69,
PL-00-681~Warszawa,~Poland\\
(b) Abteilung f\"ur Quantenphysik, Universit\"at Ulm, D-89081 Ulm, Germany}

\day{19 April 1999}

\abstract{We discuss violation of Bell inequalities by the regularized
Einstein-Podolsky-Rosen (EPR) state, which can be produced in a quantum
optical parametric down-conversion process. We propose an
experimental photodetection scheme to probe nonlocal quantum
correlations exhibited by this state. Furthermore, we show that the
correlation functions measured in two versions of the experiment are
given directly by the Wigner function and the $Q$ function of the EPR
state. Thus, the measurement of these two quasidistribution functions
yields a novel scheme for testing quantum nonlocality.}

\pacs{03.65.Bz, 42.50.Dv}

\section{Introduction}
The work of Einstein, Podolsky, and Rosen (EPR) has started a  lasting
debate about the completeness and the meaning of local realities in quantum
mechanics.  In order to make their point, EPR used the following wave function
for a system composed of two particles \cite{EPR}:  
\begin{equation}
\label{epr} 
|{\rm EPR}\rangle = \int \frac{dp}{2\pi} \ |p,-p\rangle = \int dq
\ |q,q\rangle.  
\end{equation} 
Compared to the original paper of EPR, we have put here spatial separation
between the particles equal $q_0=0$.
The above wave function describes an entangled state of two
particles $a$ and $b$, with the following properties:  
\begin{equation}
(\hat{q}_{a} -\hat{q}_{b})|{\rm EPR}\rangle=0, \ \ \ (\hat{p}_{a} +
\hat{p}_{b})|{\rm EPR}\rangle=0.  
\end{equation} 
However, further studies of quantum nonlocality and entanglement,
especially those providing quantitative tests of compatibility of quantum
mechanics with local theories in the form of Bell inequalities, used
mainly spin-1/2 particles instead of systems with continuous degrees
of freedom.

Quantum correlations for position-momentum variables associated with the
EPR state (\ref{epr}) can be analyzed in phase space using the Wigner
function or the positive $Q$ function.  Using the Wigner function
approach, Bell \cite{Bell} has argued that the EPR wave function
(\ref{epr}) will not exhibit nonlocal effects, because  its Wigner
function is positive everywhere, and as such will allow for a hidden
variable description of the correlations. This statement is correct
as long as the measured observables have a straightforward classical
interpretation in the phase space representation.  However, the situation
can change dramatically, if we take into account quantum observables
for which the phase space cannot serve as a local theory model.  In a
recent publication \cite{KBKWEPR} we have demonstrated that the Wigner
function of the EPR state, though positive definite, provides a direct
evidence of the nonlocal character of this state.  The demonstration
was based on the fact that the Wigner function can be interpreted as a
correlation function for the joint measurement of the parity operator.

In this presentation we review the EPR state in the Wigner representation
and extend our approach to the positive $Q$ function, which also can be
regarded as a correlation function for the joint measurement of certain
dichotomic observables.  The analysis will be performed for an optical
realization of the EPR state as the entangled two-mode state of light
generated in the spontaneous parametric process. In this quantum optical
context, analysis of the $Q$ function is of particular interest, as the
experimental demonstration of nonlocal correlations exhibited by the
$Q$ function poses less stringent requirements on single photon
detectors used in the setup.

This paper is organized as follows. First, in Sec.~2, we briefly review
the quantum optical version of the EPR state. In Sec.~3 we discuss
the Wigner and the $Q$ functions of this state, and study the limiting
case in which the original EPR state is recovered. In Sec.~4 we
present the quantum optical scheme which can be used to demonstrate
nonlocality of the EPR state in the phase space. Finally, Sec.~5 concludes
this presentation.

\section{EPR state in quantum optics}
Before we show that the phase space representation of the EPR state fully
exhibits the nonlocal character of this entangled state, we shall give
a brief description of an optical realization of this state, in terms
of the two-mode correlated light generated in the spontaneous parametric
down-conversion process.

With a clear application to quantum optics, it has been shown that a state
produced in a process of nondegenerate optical parametric amplification
(NOPA) \cite{Reid}, is the optical analog of the EPR state in the limit of
strong squeezing.  The NOPA process is a nonlinear interaction of
two quantized modes (denoted by $a$ and $b$) in a nonlinear medium with
a strong classical pump field. The strength of the interaction can be
characterized with the parameter $\chi$, which involves the second-order
susceptibility and the pump field amplitude.  The interaction Hamiltonian
of the system is:
\begin{equation} H= i \chi
(\hat{a}^{\dagger}\hat{b}^{\dagger} - \hat{a}\hat{b}).  
\end{equation} 
If the initial state of the
system consists of two vacuum modes the NOPA generates:  
\begin{equation}
|{\rm NOPA}\rangle = e^{r(\hat{a}^{\dagger}\hat{b}^{\dagger} - 
\hat{a}\hat{b})}
|0,0\rangle,
\end{equation} 
where $r=\chi t$ is a dimensionless parameter
characterizing the interaction time.  Simple algebra, based on the following
disentanglement of  the evolution operator:  
\begin{equation}
e^{r(\hat{a}^{\dagger}\hat{b}^{\dagger} - \hat{a}\hat{b})}
= e^{\tanh r \, \hat{a}^{\dagger} \hat{b}^{\dagger}}
\left(\frac{1}{\cosh r}
\right)^{\hat{a}^{\dagger}\hat{a} +\hat{b}^{\dagger} \hat{b} +1}
e^{-\tanh r \, \hat{a}\hat{b}},
\end{equation} 
shows that the generated state has  a diagonal
decomposition in terms of the number states of the two modes:
\begin{equation} 
\label{nopa} 
|{\rm NOPA}\rangle = \frac{1}{\cosh r} \sum_{n=0}^{\infty}
(\tanh r)^{n} |n,n\rangle.  
\end{equation} 
In order to see the relation
between this state and the EPR state (\ref{epr}), we rewrite the state
(\ref{nopa}) in the following form:  
\begin{equation} |{\rm NOPA}\rangle =
\frac{1}{\cosh r} \sum_{n=0}^{\infty} (\tanh r)^{n} \int\,dq \int\,dq'
\,|q,q'\rangle\langle q,q'| n,n\rangle.  
\end{equation} 
Using the fact that
the scalar products can be expressed in terms of the Hermite polynomials (in
dimensionless units):  
$\langle q|n\rangle\ = (2^{n}n!\sqrt{\pi})^{-1/2}
H_{n}(q)\exp(-q^{2}/2)$, and the following summation
formula (valid for arbitrary $\lambda \le 1$):  
\begin{equation} 
\label{suma}
\sum_{n=0}^{\infty} {\lambda^{n}}\langle q|n\rangle\ \langle n|q'\rangle\;
=\;\frac{1}{\sqrt{\pi(1-\lambda^{2}})}\;
\exp\left(-\frac{ q^{2}+ q'^{2} -2\lambda
qq'}{2(1-\lambda^{2})}\right), 
\end {equation} 
we obtain:  
\begin{equation}
\label{nopar}
|{\rm NOPA}\rangle = \frac{1}{\sqrt{\pi}} \int\,dq \int\,dq' 
\exp\left(-\frac{ q^{2}+ q'^{2} -2qq'\tanh r } {2(1-\tanh^2 r)}\right)
|q,q'\rangle.  
\end{equation} 
This formula is a regularized version of the
EPR state (\ref{epr}), with  a gaussian smoothing profile
of the plane waves.  Now it becomes clear, that in
the limit of $r \rightarrow \infty$ i.e., for a very long interaction time, 
this smoothing function becomes a sharp function: $\delta (q-q')$.
This follows from the fact that for $\lambda=1$, the sum (\ref{suma}) 
reduces to the completeness relation for the oscillator wave functions.  
In this limit the state (\ref{nopar}) becomes exactly the EPR state 
(\ref{epr}), and as result we obtain, that in terms of photons, 
the EPR state is:
\begin{equation} 
|{\rm EPR}\rangle \sim \lim_{r\rightarrow \infty} |{\rm
 NOPA}\rangle \sim |0,0\rangle + |1,1\rangle +|2,2\rangle + \dots \,.
\end{equation}
The NOPA state has been generated experimentally \cite{Ou} and applied to
discuss the implications of the positivity of the corresponding Wigner
function on the Bell inequality \cite{OuAPB}. 

\section{EPR state in phase space}

In this section, we discuss the Wigner and the $Q$ functions of the state
produced in the nondegenerate spontaneous parametric down-conversion
process. We shall pay particular attention to the limit of strong
interaction for $r\rightarrow\infty$, where the original, singular EPR
state is recovered.

\subsection{The Wigner function in the coherent 
state representation} 
The two mode Wigner function of the NOPA state (\ref{nopar}) can be
calculated directly from the definition:
\begin{equation}
W(\alpha;\beta) = \int \frac{ d^{2}\alpha'}{\pi^{2}} 
\int \frac{ d^{2}\beta'}{\pi^{2}} \exp(\alpha \alpha'^{\ast}-\alpha ^{\ast} 
\alpha' + \beta \beta'^{\ast}-\beta ^{\ast} \beta') 
\langle \hat{D}_a(\alpha') \hat{D}_b(\beta')\rangle\,,
\end{equation}
where $\hat{D}_a(\alpha)$ and $\hat{D}_b(\beta)$ are
the Glauber's  displacement operators
for modes $a$ and $b$.
Simple calculation shows that the Wigner function of the state (\ref{nopar})
has the form:  
\begin{equation}
\label{nopawig}
W(\alpha ;\beta) = \frac{4}{\pi^2}
\exp[-2 \cosh 2r (|\alpha|^2 + |\beta|^2) + 2 \sinh 2r (\alpha \beta +
\alpha^\ast \beta^\ast)].  
\end{equation} 
This positive everywhere Wigner
function is plotted  in Fig.~1 for real values of $\alpha$ and $\beta$ and
for $r=1$.  The Wigner function of the original
EPR state (\ref{epr}) is obtained in the limit $r\rightarrow \infty$:  
\begin{equation}
W(\alpha; \beta) \sim \delta(\alpha_{r} -\beta_{r}) \delta(\alpha_{i}
+\beta_{i}), 
\end{equation} 
where $\alpha_{r}\ (\alpha_{i})$ and $\beta_{r}\ (\beta_{i})$ are the real 
(imaginary) parts of $\alpha$ and $\beta$.  Note the
singular form of this Wigner function, due to the singular character of the
original EPR wave function (\ref{epr}).

\begin{figure}
\begin{center}
\epsfig{file=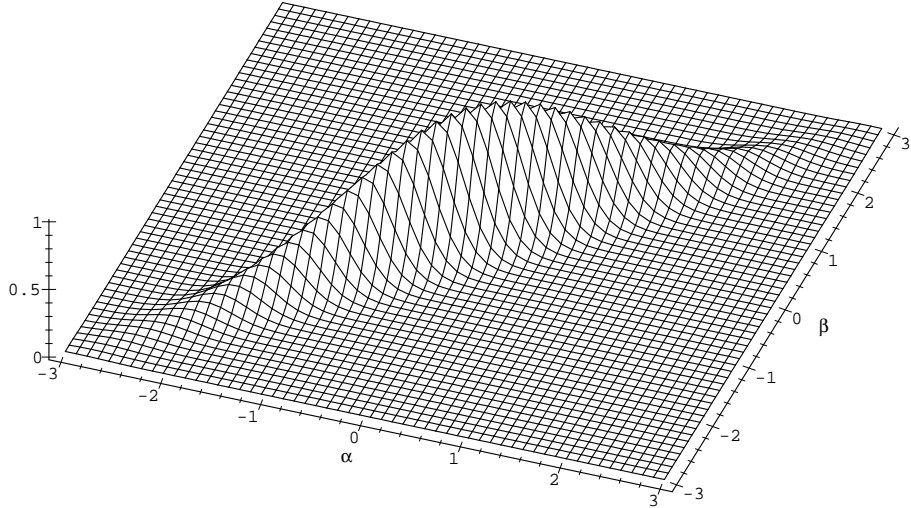,bbllx=115,bblly=78,bburx=545,bbury=690,angle=270,%
width=12cm}
\end{center}

\medskip

\caption{Fig.~1. Plot of the Wigner function of the regularized EPR
state with $r=1$ for real values of $\alpha$ and $\beta$.}

\end{figure}

\subsection{The $Q$ function in the coherent state
representation} 
The two mode $Q$ function of the NOPA state can be calculated
directly from its definition:  
\begin{equation} 
Q(\alpha; \beta)=
\frac{1}{\pi^{2}} |\langle\alpha,\beta|{\rm NOPA}\rangle|^{2}.
\end{equation} 
It is easy to show that for the NOPA state (\ref{nopar}) this function has
the following form:  
\begin{equation} 
\label{nopaQ}
Q(\alpha;\beta) = \frac{1}{(\pi \cosh
r)^{2}} \exp\left(-|\alpha|^{2} -|\beta|^{2} 
+ \tanh r (\alpha^{*} \beta^{*} +\alpha
\beta)\right), 
\end{equation} 
with the following marginals:  
\begin{eqnarray}
Q(\alpha) = \frac{1}{\pi (\cosh r)^{2}}
\exp\left(-\frac{|\alpha|^{2}} {(\cosh r)^{2}} \right)\nonumber\\
Q(\beta) = \frac{1}{\pi (\cosh r)^{2}}
\exp\left(-\frac{|\beta|^{2}} {(\cosh r)^{2}}\right).  
\end{eqnarray} 
This two-mode
$Q$ function is plotted in Fig.~2 for real values of $\alpha$ and 
$\beta$ and
for $r=1$.  
The $Q$ function of the original EPR state (\ref{epr}) obtained in the 
limit $r\rightarrow \infty$ is proportional to:  
\begin{equation} 
\label{eprQ} 
Q(\alpha; \beta) \sim
\exp(-|\alpha|^{2} - |\beta|^{2} +\alpha^{*} \beta^{*} +\alpha \beta).
\end{equation} 
The last term in the exponent is responsible for the correlation of the
entangled EPR state. Note that if we consider the $Q$ function given by
Eq.~(\ref{eprQ}) as a limit of Eq.~(\ref{nopaQ}), there is a vanishing
prefactor with the magnitude proportional to $e^{-2r}$. The function
(\ref{eprQ}) can be derived directly from the EPR wave function
(\ref{epr}):
\begin{equation}
Q(\alpha, \beta)\sim \left|\int dq \langle q,q|\alpha,\beta\rangle 
\right|^{2}.
\end{equation}
This simple integration of the coherent state wave functions, in position
representation, can be calculated and as a result we reproduce the result 
(\ref{eprQ}).

\begin{figure}
\begin{center}
\epsfig{file=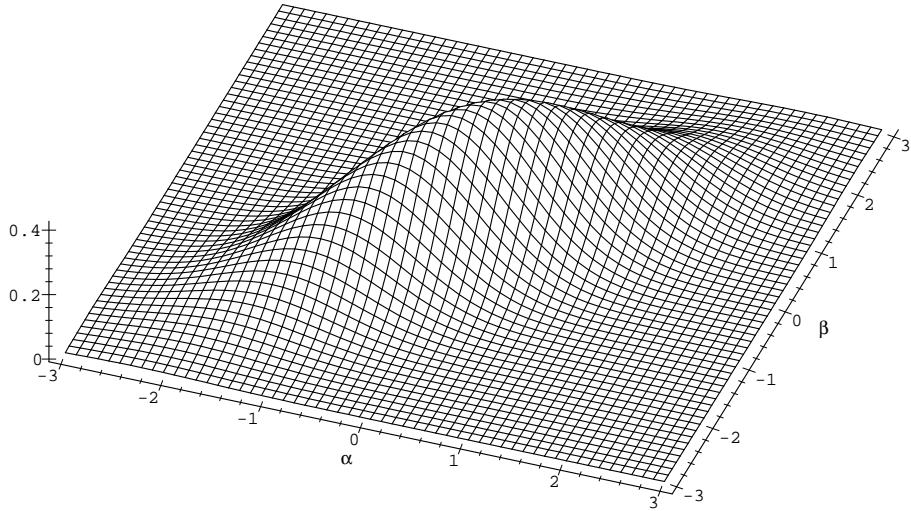,bbllx=115,bblly=78,bburx=545,bbury=690,angle=270,%
width=12cm}
\end{center}

\medskip

\caption{Fig.~2. Plot of the $Q$ function of the regularized EPR
state for real values of $\alpha$ and $\beta$, for $r=1$.}
\end{figure}

\section{Testing quantum nonlocality in phase space}
In this presentation we propose to probe the correlations of the NOPA
state with a scheme involving two photon counting detectors for modes
$a$ and $b$. The setup to demonstrate quantum nonlocality for the Wigner
function and the $Q$ function is presented in Fig.~3.  The photons from
the NOPA source are  measured by two photon counting detectors preceded
by two beam splitters.  These beam splitters, with the transmission
coefficients very close to one, mix the NOPA photons with two highly
excited coherent states.  It has been shown elsewhere \cite{KBKWPRL96},
that in the limit of the transmission tending to one, the effect of
the beam splitters is equivalent to phase space shifts for the two
modes by $\hat{D}_a(\alpha)$ and $\hat{D}_b(\beta)$, where $\alpha$
and $\beta$ are complex amplitudes of the reflected coherent fields.
A single-mode version of such an experiment, involving measurements of
the displaced photon number statistics for simple classical states of
light, is reported in these proceedings in the context of quantum state
measurement \cite{ExpWig}.

We shall show that if the detectors placed in the two arms of the
proposed measurement scheme can resolve the number of absorbed photons,
the phase space Wigner function can be determined directly. Moreover,
it will become apparent that the Wigner function measured in this setup
describes correlations between the parity of the number of photons
registered by the two detectors. As the parity is a dichotomic $\pm
1$ variable, the Wigner function can be therefore directly inserted into
appropriate Bell inequalities in order to test the nonlocal character
of the NOPA state.

The most efficient single-photon detectors available currently are
avalanche photodiodes operated in the Geiger regime. These detectors
cannot resolve the number of simultaneously absorbed photons, and provide
only a binary answer saying whether any photons have been registered
or not \cite{Kwiat}. We shall show that use of these detectors in our
scheme leads in a natural way to the measurement of the two-mode $Q$
function, and that the binary answer of the detectors makes the $Q$
function a nonlocal correlation function which violates the corresponding
Bell inequality.

\begin{figure}
\begin{center}
\epsfig{file=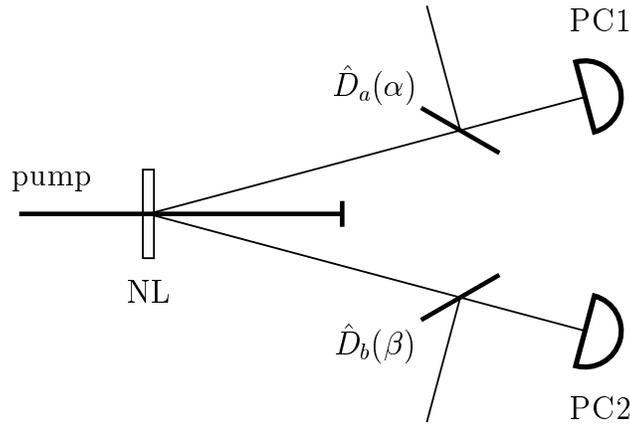,bbllx=130,bblly=340,bburx=460,bbury=540,width=10cm}
\end{center}

\medskip

\caption{Fig.~3. The experimental scheme proposed to test nonlocality
exhibited by the quantum optical realisation of the EPR state. The
nonlinear crystal NL pumped by the coherent field generates two correlated
beams. The high-transmission beam splitters with auxiliary coherent
fields placed in the paths of these beams perform the displacement
transformations $\hat{D}_a(\alpha)$ and $\hat{D}_b(\beta)$. The phase
space displaced beams are monitored by the photon counting detectors
PC1 and PC2.}
\end{figure}

\subsection{Violation of local reality by the
Wigner function}

In order to  relate  the Wigner function to photocount correlations,
we will consider the case when the detectors are capable of
resolving the number of absorbed photons. Let us assign to each
event $+1$ or $-1$, depending on whether an even or an odd number
of photons has been registered.
In this  case the joint correlation for the photon number parities is:
\begin{equation} 
E(\alpha ;\beta) = \langle{\rm NOPA}|
\hat{\Pi}_{a}(\alpha) \otimes \hat{\Pi}_{b}(\beta) |{\rm NOPA}\rangle,
\end{equation}
where 
\begin{equation} 
\hat{\Pi}_{a}(\alpha) = \hat{D}_a(\alpha)
(-1)^{\hat{n}_a} \hat{D}^{\dagger}_a(\alpha), 
\ \ \ \ \ \hat{\Pi}_{b}(\beta) = \hat{D}_b(\beta)
(-1)^{\hat{n}_b} \hat{D}^{\dagger}_b(\beta), 
\end{equation}
are parity operators of the modes $a$ and $b$ characterized by the
corresponding photon number operators $\hat{n}_a $ and $\hat{n}_b $.
Equivalently, $E(\alpha ;\beta)$ can be written as the overall parity
operator displaced in the two-mode Hilbert space by the operation
$\hat{D}^{\dagger}_a(\alpha) \otimes \hat{D}^{\dagger}_b(\beta)$.  This
observable is known in the phase space representation to provide,
up to the normalization constant, the
Wigner function of the quantum state \cite{KBKWPRL96}.
On the other hand, we can establish a clear
analogy of this scheme with joint spin-1/2 measurements. The
counterpart of the polarizer orientation is now the coherent
displacement, which can be set freely in each of the two spatially
separated apparatuses. Furthermore, classification of the registered
number of photons according to the parity provides a dichotomic $\pm 1$
outcome which is analogous to the spin direction.  As a result all
types of Bell's inequalities derived for spin systems can be used to
test the nonlocality of the NOPA Wigner function.  In Ref.~\cite{KBKWEPR}
we have shown that the Wigner function
(\ref{nopawig}) leads to a violation of the the Bell inequality. Here we
will just quote the result. For local hidden variable theories the
following  combination:
\begin{equation}
{\cal B} = E(\alpha';\beta') + E(\alpha';\beta)+E(\alpha;\beta')-
E(\alpha;\beta)
\end{equation}
should satisfy the Bell inequality $-2\le {\cal B} \le 2$. We have found that 
for a certain selection of coherent displacements the value of this
combination in the limit $r\rightarrow \infty$ is: 
${\cal B} = 1 + 3 \cdot 2^{-4/3} \approx 2.19$, which
clearly violates the Bell inequality. 

\subsection{Violation of local reality by the $Q$ function} 
Based on the previous result we shall demonstrate how nonlocality of
the EPR state can be revealed using the positive definite $Q$
quasidistribution function. The quantum optical experiment we shall
propose does not require detectors that can resolve the number of
photons triggering the output signal.
We will be interested in events when {\em no photons} were registered.
This measurement is described by the following observable:
\begin{equation}
\hat{O}_{a}=\lim_{\epsilon\rightarrow 0} \left( \epsilon\right)^{\hat{n}_{a}}
= \; : \! e^{-\hat{n}_{a}} \! : \; = |0_{a}\rangle\langle 0_{a}|, 
\end{equation} 
for the mode $a$, and a similar observable $\hat{O}_{b}$ for the mode $b$.  
As we discussed in the previous subsection, these observables 
can be shifted in phase space using auxiliary coherent
fields and high-transmission beam splitters.
As a result of this shifting we obtain:
\begin{equation} 
\hat{O}_{a}(\alpha) = \hat{D}_a(\alpha) \hat{O}_{a}
\hat{D}^{\dagger}_a(\alpha), \ \ \ 
\hat{O}_{b}(\beta) = \hat{D}_b(\beta) \hat{O}_{b}
\hat{D}^{\dagger}_b(\beta). 
\end{equation}
Obviously, these observables describe projections on a coherent state
$|\alpha\rangle\langle\alpha|$ for the mode $a$ and 
$|\beta\rangle\langle\beta|$ for the mode $b$. Therefore, statistics
of no-count events yields directly the $Q$ function of the measured field.

\begin{figure}
\begin{center}
\epsfig{file=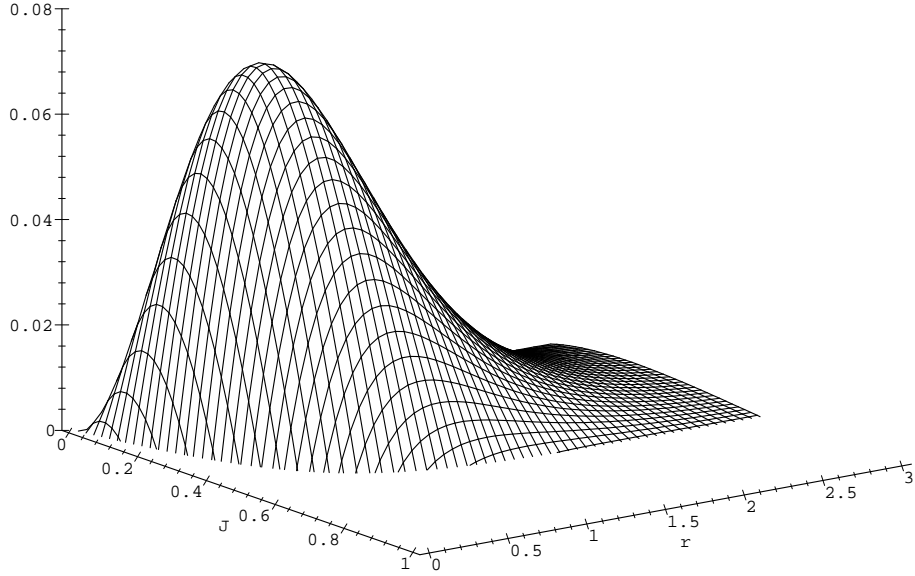,bbllx=115,bblly=78,bburx=545,bbury=690,angle=270,%
width=12cm}
\end{center}

\medskip

\caption{Fig.~4. Plot of the combination ${\cal CH}$ as a function
of the displacement intensity $J$ and the interaction parameter $r$.
Only values exceeding the upper bound imposed on ${\cal CH}$
by local theories are shown.}
\end{figure}

The quantum mechanical probability of no-count
events in both the detectors is:  
\begin{eqnarray} 
p_{ab}(\alpha;\beta) & = &
\langle{\rm NOPA}| \hat{O}_{a}(\alpha) \otimes \hat{O}_{b}(\beta) |{\rm
NOPA}\rangle \nonumber \\ & =& \frac{1}{ (\cosh r)^{2}} 
\exp\left( -|\alpha|^{2}
-|\beta|^{2} + \tanh r (\alpha^{*} \beta^{*} +\alpha \beta)\right)
\end{eqnarray}
where $\alpha$ and $\beta$ are the coherent displacements for the modes
$a$ and $b$, respectively.  The probabilities of no-count events on
single detectors are:
\begin{eqnarray} 
p_{a}(\alpha) & = & \langle {\rm NOPA} | \hat{Q}_{a}(\alpha)
\otimes \hat{I}_{b} |{\rm NOPA} \rangle = \frac{1}{ (\cosh r)^{2}}
\exp\left(-\frac{|\alpha|^{2}} {(\cosh r)^{2}}\right),
\nonumber 
\\ p_{b}(\beta) & = &
\langle {\rm NOPA} | \hat{I}_{a} \otimes \hat{O}_{b}(\beta) | {\rm
NOPA}\rangle = \frac{1}{ (\cosh r)^{2}} 
\exp\left(-\frac{|\beta|^{2}} {(\cosh
r)^{2}}\right).  
\end{eqnarray} 
We recognize that these probabilities are equal up to a normalization
factor the two-mode $Q$ function (\ref{nopaQ}) and its two marginals
for the NOPA state.

The correlation function $p_{ab}(\alpha,\beta)$ describes a binary $0$ or $1$
measurement on the modes $a$ and $b$ with adjustable parameters of the
apparatuses represented by $\alpha$ and $\beta$.
As a result of this, 
the Bell inequality
derived for a measurement of local realities bounded by $0$ and $1$,
can be applied to test the nonlocal character of the NOPA state
(\ref{nopa}), using the $Q$ function.  We shall consider the the
Clauser-Horne combination \cite{CH} for a selected set of four displacements:
\begin{equation}  
{\cal CH} = p_{ab}(0;0)+ p_{ab}(\alpha;0) +
p_{ab}(0;\beta) - p_{ab}(\alpha;\beta)- p_{a}(0) - p_{b}(0), 
\end{equation}
which for local theories satisfies the inequality $-1 \le {\cal CH} \le 0$.
We will take the coherent displacements to  be real with $\alpha = -\beta
=\sqrt{J}$.  For these values we obtain 
\begin{equation}
 {\cal CH} = \frac{1}{ (\cosh r)^{2}}\left(2 e^{-J}
-e^{-2 J(1+ \tanh r)} -1\right).  
\end{equation}
As depicted in Fig.~4, this result violates the upper bound imposed by local
theories. In the limit of $r \rightarrow \infty$ and small values
of the displacement, this function becomes:
\begin{equation}
{\cal CH}\approx 8J e^{-2r},
\end{equation}
i.e., it violates the Bell inequality for all values of $r$, but
the violation becomes
infinitesimally small. This results from the fact that the $Q$ function
(\ref{eprQ}) in this limit is dumped by the prefactor
$1/\cosh^2 r$, decreasing like $e^{-2r}$.

\section{Conclusions}
In this presentation we have shown that nonlocality of the EPR state
can be revealed using its phase space representation of the form of the
Wigner or $Q$ quasidistribution functions. This is possible owing
to the observation that these quasidistribution functions describe
nonlocal correlations that can be detected in a certain quantum optical
scheme. Of course, this scheme is much more general. It can be applied
to measure an arbitrary two-mode state of light, and the corresponding
Wigner and $Q$ functions will always have the operational meaning of
nonlocal correlation functions \cite{KBKWPRL99}.

\section*{Acknowledgements} 
This work has been partially supported by the Polish KBN grant No.
2 P03B 089 16. K.W. thanks the Alexander von Humboldt Foundation for
generous support and Prof.\ W.~P.~Schleich for his hospitality in Ulm.

\end{document}